# Timescales and contribution of heating and helicity effect in helicity-dependent all-optical switching

Guanqi Li[1,2], Xiangyu Zheng[1,2], Junlin Wang[1,2], Xianyang Lu[2], Jing Wu[1, 2] *, Jianwang Cai[3]*, Thomas A. Ostler[4], and Yongbing Xu[2, 5, *]

[1]Department of Physics, University of York, York, YO10 5DD, UK

[2]York-Nanjing International Joint Center in Spintronics, School of Electronic Science and Engineering, Nanjing University, Nanjing, 210093, China

[3]Institute of Physics, Chinese Academy of Sciences, Beijing 100190, China

[4]College of Business, Technology and Engineering, Sheffield Hallam University, Sheffield, S1 1WB, UK

[5]Spintronics and Nanodevice Laboratory, Department of Electronics Engineering, University of York, York, YO10 5DD, UK

*Corresponding author. E-mail: jing.wu@york.ac.uk, jwcai@aphy.iphy.ac.cn, yongbing.xu@york.ac.uk

Abstract

The manipulation of the magnetic direction by using the ultrafast laser pulse is attractive for its great advantages in terms of speed and energy efficiency for information storage applications. However, the heating and helicity effects induced by circularly polarized laser excitation are entangled in the helicity-dependent all-optical switching (HD-AOS), which hinders the understanding of magnetization dynamics involved. Here, by applying a dual-pump laser excitation, first with a linearly polarized (LP) laser pulse followed by a circularly polarized (CP) laser pulse, we identify the timescales and contribution from heating and helicity effects in HD-AOS with a Pt /Co /Pt triple layer. When the sample is preheated by the LP laser pulses to a nearly fully demagnetized state, CP laser pulses with a much-reduced power switches the sample's magnetization. By varying the time delay between the two pump pulses, we show that the helicity effect, which gives rise to the deterministic helicity induced switching, onsets instantly upon laser excitation, and only exists for less than 0.2 ps close to the laser pulse duration of 0.15 ps. The results reveal that the transient magnetization state upon which CP laser pulses impinge is the key factor for achieving HD-AOS, and importantly, the tunability between heating and helicity effects with the unique dual-pump laser excitation approach will enable HD-AOS in a wide range of magnetic material systems for the potential ultrafast spintronics applications.

## INTRODUCTION

The all-optical magnetization switching (AOS) was first observed in a landmark experiment (*1*) demonstrating that magnetization can be reversed by laser pulses without any applied magnetic field in GdFeCo. The microscopic mechanism for the AOS process in GdFeCo has been considered to be a helicity-independent heating effect, which is due to the fact that the Fe and Gd sublattices demagnetize on very different timescales (*2*). This leads to a transient ferromagnetic-like state discovered in GdFeCo, which mediates the helicity-independent all-optical switching (HID-AOS) (*3*). HID-AOS is a single-pulse thermal switching (*4, 5*). It is not limited to Gd-based ferrimagnetic alloys but also in Gd-based ferrimagnetic multilayers (*6, 7*) and ferrimagnetic Heusler alloy $Mn_2Ru_xGa$ (*8*). However, AOS is also found to be helicity-dependent in some materials such as ferrimagnetic Tb-transition metal (TM) alloys (*9*), synthetic ferrimagnets (*10*), and ferromagnetic Co/Pt multilayers (*11*). It was believed initially that the circularly polarized light simultaneously acts as an effective magnetic field, due to the inverse Faraday effect (IFE), in the spin system during helicity-dependent AOS (HD-AOS) (*12-14*). The strength and the lifetime of the induced field pulses are still a matter of debate. When HD-AOS is studied in the ultrafast time domain (*9, 15*), the effective magnetic field strength due to IFE has to be as large as 10 Tesla to achieve such a short switching time according to the theoretical simulations. A minimal IFE lifetime of 0.15 ps was estimated for Co/Pt (*12*) while longer duration time up to several ps after the laser pulse excitation were also reported (*16-19*). The IFE response has been found to be strongly material dependent, and the interlayer exchange interactions and spin-orbit coupling are considered to play an important role in HD-AOS of magnetic multilayers (*20*). An optical spin-transfer torque was also suggested to play a role in HD-AOS of ferromagnetic thin films with a Pt capping layer (*21*). Other mechanisms put forward to account for HD-AOS include the laser-induced heating (*22, 23*), magnetic circular dichroism (MCD) (*24, 25*), and optical selection rule (*26*).

Heating and helicity effects are entangled in HD-AOS using circular polarized lasers, and the individual contribution cannot be distinguished by using just one circularly polarized pump beam. Furthermore, as the HD-AOS has been reported as a multi-pulse effect (*27*), the important information, such as the onset and duration of helicity effects, and interplay between the heating and helicity effects in the first few picoseconds of HD-AOS switching processes cannot be obtained from a conventional time-resolved measurement with a single pump beam (*11, 15, 19, 24*). In this paper, we have employed a dual-pumping scheme combined with a magneto-optical microscopic detection to investigate HD-AOS in a Pt/Co/Pt triple-layer sample, as shown schematically in FIG. 1(a), to identify individual contributions from heating and helicity effects, and distinguish the time dependence between them.

This novel dual-pumping scheme allows us to choose the helicity of each pump pulse independently and vary both the power combination and time delay between the two pump pulses. The transient pre-heated state of the electron-spin system, on which the second CP pump pulse is to impinge, can be tuned by controlling of the power of the first LP pump pulse as well as the precise time delay between the two pulses, as indicated in FIG. 1(b). We have found that HD-AOS can be achieved with a circularly polarized laser pulse of very low fluencies, if a linearly polarized laser (LP) pulse is used to preheat magnetization. The strong correlation between HD-AOS and the time interval between the LP and CP pulses signposts an instant onset of helicity effect, but only lasting for a period of the order of the laser pulse duration. The pre-heated transient demagnetization state is found to be a key factor for realizing HD-AOS.

## RESULTS

HD-AOS induced by a sequence of 100 LP and CP pulse pairs with different power combinations and a fixed delay time, 1.6 picoseconds, between them was investigated first. At this delay time, the hot electrons/spins excited by the first LP pulse has reached their thermal equilibrium state with the system temperature close to its peak value when the second CP pulse arrived. Fig. 2 shows each subtracted MOKE image centered along the scanning path of the dual-pump beam with a field of view of 60 μm × 380 μm. The horizontal axis shows the total power of the LP and CP pulse pairs increasing from 100 μW to 180 μW, while the vertical axis gives the power of the CP pulse increasing from 10 μW to the the total power of each column. The power of the LP pulse used for each image is the difference between the total power and the CP pulse power. Therefore, the power of the LP pump pulse is zero at the bottom of each column and the corresponding magnetic domain images, framed in green dashed line, are induced by the CP pulses only. As shown in the first column of Fig. 2, there is no laser pumping induced change in the sample's magnetic state except of the 90 µW LP + 10 µW CP when the total power is 100 µW. The images of the first row show that a random domain state was induced when the CP beam power is only 10 µW. Once the CP beam power is increased to 20 µW and beyond, HD-AOS was observed with a total power window of 120 µW ($10.6$ mJ/cm$^2$) to 160 µW ($14.1$ mJ/cm$^2$). It indicates that the power window, in which HD-AOS occurs, for pairs of LP and CP pulses is the same as that for a single CP pulse. This shows that the energy dissipation from the heated electron/spin system is negligible when the second CP arrives at a 1.6 ps time delay, which leads to the same HD-AOS power window (120 µW - 160 µW) as that in a single CP pump excitation. As shown in the images framed in dashed red lines in Fig. 2, the laser-swept area remains a uniformly switched magnetic domain even when the power of the CP beam is reduced to 20 µW with the samples preheated with the LP pulse. When the CP beam power is less than 20 µW, no HD-AOS was observed. Therefore, the minimum power of the CP pulse required to achieve HD-AOS is 20 µW when the sample is preheated. It gives a threshold laser fluence for helicity effect as low as 1.77 mJ/cm$^2$, only 20% of the total laser fluence 10.6 mJ/cm$^2$ (120 µW), which would be the apparent threshold value in a single CP pump measurement as shown by the images framed with green dashed lines. This proves that the laser heating plays an essential role in HD-AOS of the Pt/Co/Pt triple layer, where only a single magnetic lattice exists, in comparison with HID-AOS in RE-FM alloy/multilayers [5,6]. When the spin temperature is high enough, circularly polarized illumination with a power threshold as low as 1.77 mJ/cm$^2$ is sufficient to achieve HD-AOS as demonstrated by the images framed in red dashed lines in Fig. 2. Without pre-heating by a LP pulse, the CP illumination has to be about five times as intense in order to trigger HD-AOS. This discovery reveals that in a single-pump-induced HD-AOS event in Pt/Co/Pt material [20], most of the required pulse energy is used to heat spin system. The helicity effect requires only a small portion of the power threshold. In the column of the 160 µW total power, multidomain patterns started to emerge in the center of the laser beam path. When the total power was increased to 180 µW, multidomain patterns were induced regardless of power combination. The occurring of the multidomain states under high power laser pumping is due to the laser overheating which demagnetizes the sample again after HD-AOS.

As indicated in Fig. 1(b), the delay time between the LP and CP on HD-AOS is a critical factor. To study this effect in detail, the delay time was set from 0 to 10 ps, with a step size of 0.2 ps for the first 2 ps, and then 0.5 ps afterwards. The CP beam power was increased from 20 to 100 µW with a step size of 10 µW, while the LP was decreased from 100 to 20 µW, so that the total power was fixed at 120 µW, which is the minimum total laser power needed for HD-AOS. The switching ratio of HD-AOS was extracted for each MOKE image captured at every delay time, quantitated via image processing using ImageJ (*28*), and plotted as a function of time delay in Fig S5, 6 (see supplementary information for details). The MOKE images and the extracted switching ratio of two representative power combinations are displayed in Fig. 3. The interference of the two pump pulses at zero delay point induced a multidomain state, which leads to an approximate 50% switching ratio for every curve in Fig. 3(c), (d) at zero delay point. With the increase of the time delay, the switching ratio increases first and reaches its highest point, approximately 90%, when the time delay is about 1 ps for all the power combinations. However, after the initial rise, the switching rate shows a significant difference of its dependence on the LP and CP time delay between these two power combinations. For the case of LP power 40 µW ($3.53$ mJ/cm$^2$) and CP power 80 µW, the switching ratio drops sharply when the time delay between the two pulses are longer than 2 ps. It decreases to less than 20% when the time delay is longer than 3 ps as shown in Fig. 3(c). On the other hand, for the case of LP power 80 $\mu W$ ($7.06$ mJ/cm$^2$) and CP power 40 µW, the switching ratio stays at its highest value (~90%) for the time delay from 1 ps to 4 ps. When the time delay is longer than 4 ps, the switching ratio only drops to around 75%, and stays at this high rate over a time delay window as long as 100 ps. Then it drops very slowly to 60% over the next 300 ps and fast afterwards as shown in Fig. 3(d). These two different processes are also evidenced in their MOKE images at different time delay as presented in FIG. 3(a) and (b), where FIG. 3(a), for the case of LP power 40 µW, shows no sign of switching at 6 ps, while FIG. 3(b) a clear switching at the same delay time, but with a larger LP power of 80 µW.

A two-temperature model has been employed to simulate the demagnetization rate and magnetization recovery after the LP pulse for both cases, i.e. of 40 and 80 µW, and the results are superimposed on their switching rate curves in Fig. 3 (c) and (d), respectively. It shows that the magnetization recovery (spin cooling) time after laser excitation increases from a couple of picoseconds to nanosecond time scales with a moderate increase of the excitation power, which was also observed in previous experiments and simulations (*7, 29, 30*). A red dotted horizontal line is drawn at 50% switching ratio (left hand y axis) in both Fig. 3 (c) and (d). In Fig. 3 (c), HD-AOS occurs when the 80 µW CP pulse arrives when the sample magnetization recover to less than ~70% of its saturation value. In Fig. 3 (d), HD-AOS occurs when the 40 µW CP pulse arrives when the sample magnetization recover to less than ~60% of its saturation value. This further indicates that the demagnetization state upon which the CP pulse impinges is a key factor to achieve HD-AOS. Furthermore, the required demagnetizing degree is dependent on the power of the CP pulse. A higher degree of demagnetization state requires a lower CP pulse to achieve HD-AOS. However, the CP pulses also heats up the sample magnetization. The 80 µW CP pulse in Fig. 3 (c) should re-demagnetize the sample to a degree as high as, if not higher than, that caused by the 80 µW LP pulse in Fig. 3 (d). With a stronger helicity effect from the 80 µW CP pulse, HD-AOS in the case of 40 µW LP + 80 µW CP was expected to occur in a similar time interval range of, if not even longer than, in the case of 80 µW LP + 40 µW CP. The essential role of heating in HD-AOS on its own cannot explain the dramatically different time-delay dependences between these two cases. This difference suggests that the action of the helicity effect comes to an end before the spin temperature reaches its second peak caused by the CP pulses.

**DISCUSSION**

To make this picture clear, the two-temperature model has been applied again to simulate the demagnetization rate and magnetization recovering excited by both the LP and CP pulses. Fig. 4 (a) and (b) show the case when two pulses are 5 ps apart. The spin flip energy barriers related to the spin temperatures are also added and represented by $E_{fa}^t$ and $E_{fb}^t$, respectively. To identify the lag between helicity effect and heating effect, we show the corresponding $E_f^t$ at four time points: the first pulse arrived time ($t = 0$), the first demagnetization peak ($t = 1\ ps$), the second pulse arrived point ($t = 5\ ps$), and demagnetization peak ($t = 6\ ps$). One can see that in case of Fig. 4 (b) with the LP 80 µW + CP 40 µW pulse combination, at 5 ps delay time, the transient magnetization is still much lower than 60% of its saturation value, the required demagnetizing degree for a 40 µW CP pulse. This corresponds to a lower energy barrier $E_{fb}^5$ between $M^\uparrow$ and $M^\downarrow$ states upon the arrival of the 40 µW CP pulse, and HD-AOS takes place in this case. In the case of Fig. 4. (a) with 40 µW+ 80 µW pulse pair, at 5 ps delay time, the transient magnetization recovers to around 80% of its saturation value, higher than the required demagnetizing degree (70% of the saturation value) for an 80 µW CP pulse. This corresponds to a high energy barrier $E_{fa}^5$ between $M^\uparrow$ and $M^\downarrow$ states upon the arrival of the 80 µW CP pulse. Even though the transient magnetization is reduced to about 70% of its saturation value after another 0.2 ps time delay, as marked by two short vertical dash lines (blue) in Fig. 4 (b), HD-AOS doesn't take places as the laser's heating effect takes more than 0.3 ps to reach a high demagnetization degree. The only explanation for this observation is that the onset time of helicity effect from the CP pulse is instant, and the duration of the helicity effect is less than 0.2 ps close to the laser pulse width of 150 fs. Even though the energy barrier $E_{fa}^6$ is reduced further by the heating effect of the CP pulse itself, the helicity effect has already disappeared at this point, and HD-AOS cannot be triggered anymore. The LP pulse heating induced demagnetization has two sides. It assists helicity effect in achieving HD-AOS, upon which CP pulses impinge. Once the helicity effect ends, it starts to demagnetize reversed magnetization from the helicity effect.

We have further investigated the relationship between HD-AOS switching ratio and laser ellipticity using a single pump and the results are included in Fig. S4. The switching ratio was found to decrease as the laser polarization changes from circular to linear. This is consistent with a previous finding on laser-induced domain wall motion where wall displacement decreases as laser polarization changes from circular to linear (*15*). For these single pulse cases, the LP and CP photons arrived at the same time, and the heating effect from the LP photons lags and thus the CP photons fail to achieve HD-AOS. This might be the reason that HD-AOS has not been observed in a wider range of material systems, because, generally, ultrafast laser heating effects lag behind its helicity effect. This also explains a previous observation that a longer laser pulse duration gives a higher switching ratio under the same laser power AOS (*31*). As shown in Fig. S8, the higher laser fluence takes a longer time to reach the highest demagnetization state as pointed out previously (*30*). With the dual-pump laser pulses, we expect that HD-AOS would occur in many other magnetic materials where the transient magnetization states needed for the CP lighted driven HD-AOS can be achieved by controlling the strength of the LP pulse and the delay time.

In conclusion, we have applied dual-pulse laser excitation to identify the contribution and time dependence of heating and helicity effect in HD-AOS in a Pt/Co/Pt triple layer. We have shown that pre-heating plays an essential role in HD-AOS. The laser power requested by helicity effect in HD-AOS could be very low when magnetization is close to a fully demagnetized state. By varying the time delay between LP and CP pulses with different energy combinations, we have demonstrated unambiguously that the helicity effect, which gives rise to the deterministic helicity-induced switching, occurs instantly upon laser excitation, and only exists over the laser pulse duration. This work has disentangled the heating and helicity effects and revealed their timescales in helicity-dependent all-optical magnetization switching. At the same time, the unique LP/CP dual-pump scheme makes the manipulation of HD-AOS feasible, which provides a promising way for achieving HD-AOS in a wide range of material systems.

## MATERIALS AND METHODS

Sample fabrication

The sample Pt (2 nm)/Co (0.6 nm)/Pt (2 nm) triple layer was fabricated on a Corning glass substrate by magnetic sputtering. The base pressure for the sputtering system was better than $4 \times 10^{-5}$ Pa. The working Ar pressure was 0.5 Pa. The thickness of the Corning glass substrate is 0.13 mm. A 5 nm Ta buffer layer was deposited prior to the Pt/Co/Pt growth, which improves the Pt/Co interface smoothness and the (111) orientation, and therefore enhance the perpendicular magnetic anisotropy of the Co layer, as confirmed in the MOKE hysteresis loop shown in Fig. S1.

Experimental Method

A Ti:sapphire laser amplifier system with 150 fs pulse duration and 800 nm central wavelength was used. For the dual-pumping measurements, the pulse was split into two pulses. The first pump pulse was linearly polarized (LP) and used to heat up the sample's electron/spin systems. The second pump pulse was circularly polarized (CP), delayed with respect to the first LP pulse, and used to switch the sample's magnetic state, as illustrated in Fig. 1(a). The power of each pump beam was individually adjusted for a desired power combination. The two pump beams were made co-linear before being focused onto the Pt/Co/Pt triple layer from the substrate side. The spot size was measured as 38 μm in diameter using a CCD beam profiler, which gave a laser fluence of $8.83 \times 10^{-2}$ mJ/cm$^2$ for laser power at 1 μW. The sample was mounted on a motorized 3-axis nanomax flexure stage. The magnetization of the sample was initially saturated along the perpendicular direction of the sample plane defined as $M^\uparrow$ state. When the sample was exposed to the dual-pump beams, the stage was scanned over a 300 μm distance at a velocity of 10 μm/s. It is equivalent to 100 pulses' excitation per every 1 μm illuminated path on the sample from each pump beam. After laser excitation, the magnetic domain state was recorded as a MOKE image via a wide-field Magneto-Optical Kerr (MOKE) microscope. Then the sample was re-magnetized to $M^\uparrow$ state and a reference MOKE image was taken. The MOKE images presented are the subtractions of each pair of these images, where any effects from the surface morphology are eliminated.

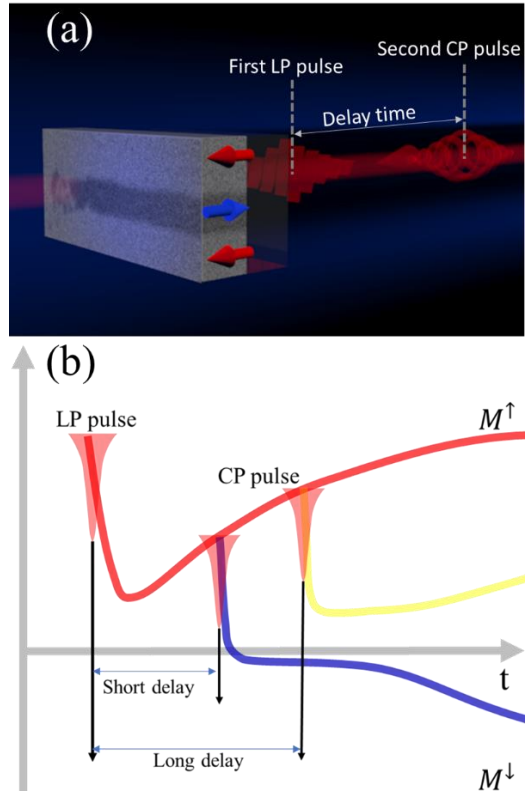

FIG. 1. Double-pump scheme and its implementation. (a) A magnetic domain image of the Pt/Co/Pt triple-layer sample under laser illumination from the substrate side. The magnetization is initially saturated along the perpendicular direction of the sample plane labeled as $M^\uparrow$ state (red arrows). The area exposed under laser is switched to the opposite direction labelled as $M^\downarrow$ state (blue arrow). (b) The magnetization of the exposed area as a function of time with dual-pulse excitation. The first linearly polarized (LP) pulse heats the sample to a demagnetization state (red curve). The second circularly polarized (CP) pulse arrives after a certain delay. For a short delay, the domain switching is expected (blue curve). But for a long delay, the switching may not occur (yellow curve).

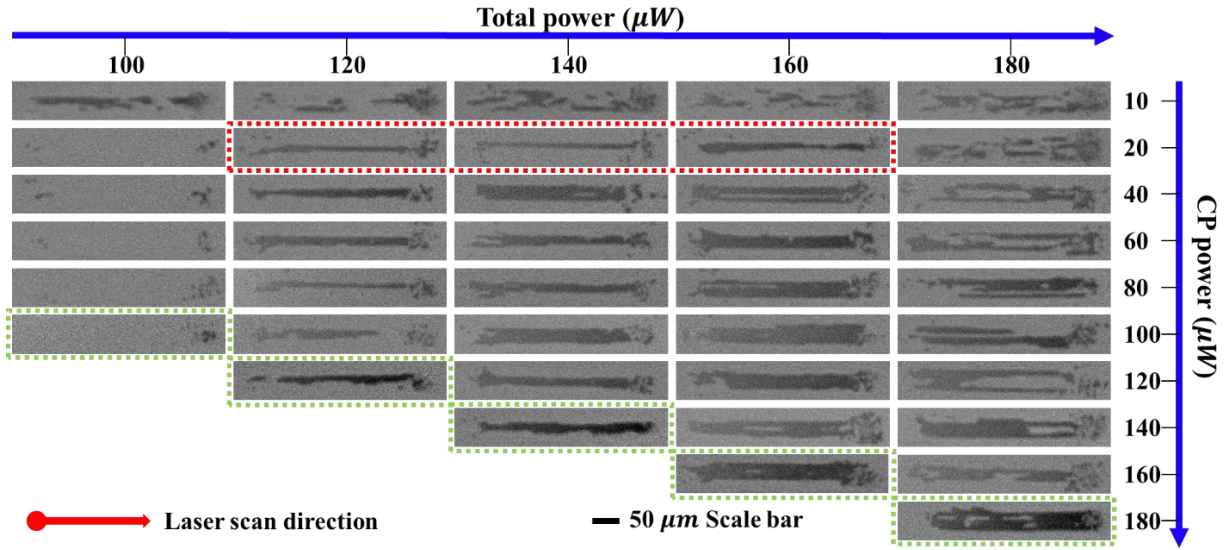

FIG. 2. HD-AOS induced by dual-pump with different power combinations at a fixed time interval. MOKE images of magnetic domains induced by a sequence of 100 LP and CP pulse pairs with different power combinations and a fixed delay time, 1.6 ps, in a Pt/Co/Pt triple layer. The horizontal axis shows the total power of the LP and CP pulse pairs, increasing from 100 μW to 180 μW, while the vertical axis gives the power of the CP pulse increasing from 10 μW to the the total power of each column. The green frame highlights the images with CP pulses alone showing a threshold CP power of 120 μW needed for the HD-AOS effect, while the red frame highlights the images showing clear HD-AOS effect with the CP power as low as 20 μW after preheating with LP light.

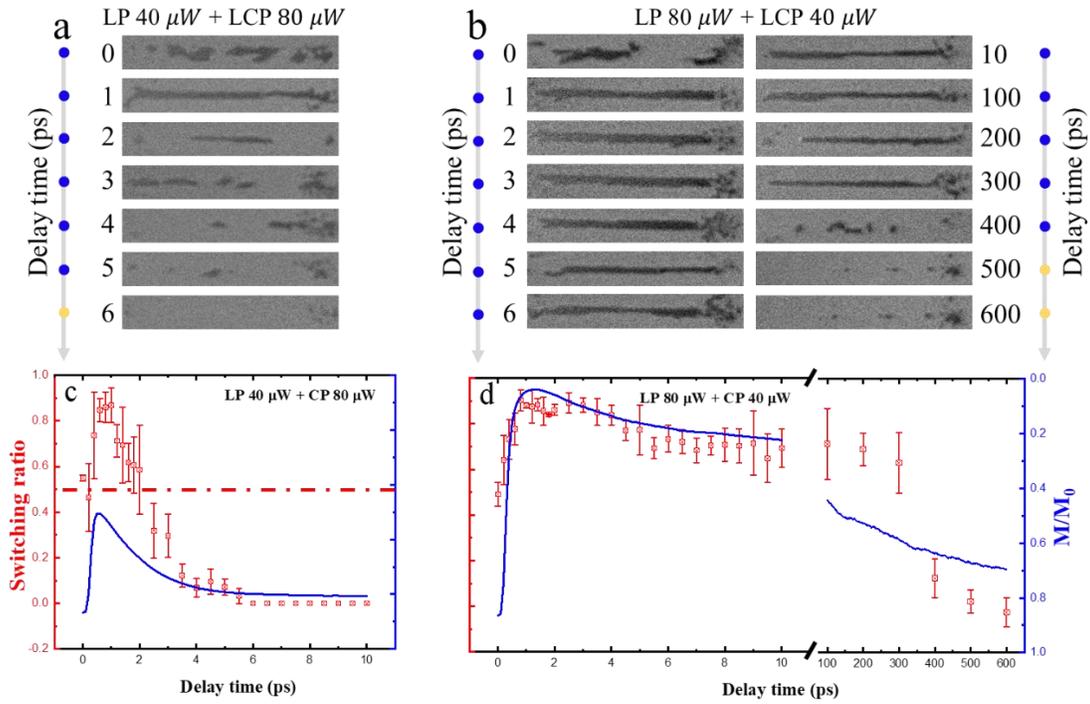

FIG. 3. HD-AOS induced by dual-pump as a function of time intervals. The effect of the delay time between the LP and CP pulses on HD-AOS in a Pt/Co/Pt triple layer. MOKE images of magnetic domains induced under two different combinations of LP and CP powers (a) LP 40 μW + CP 80 μW and (b) LP 80 μW + CP 40 μW under the same total power of 120 μW. The number next to each image indicates the delay time. The corresponding switching ratio vs delay time are shown in (c) and (d) in red square symbol. The horizontal red-dotted line indicates a switching ratio of 50%. The superimposed blue lines are the simulated demagnetization curves from the LP pump excitation only, indicating the transient magnetization state before the arrival of the CP pulse.

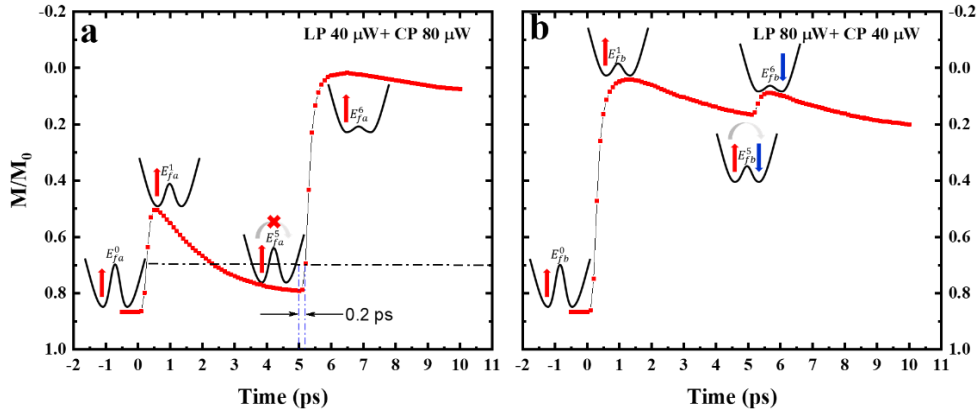

FIG. 4. Simulation of dual-pump induced magnetic switching process. The temporal profiles of the simulated demagnetization induced by both LP and CP pulses at 5 ps time delay for two power combinations, (a) 40 μW+ 80 μW and (b) 80 μW+ 40 μW. Two short vertical dashed lines (in blue) in (a) mark the time when the magnetization crosses the 70% line (black dashed line) after CP pulse illumination. The effective energy barriers ($E_{fa}^t$, $E_{fb}^t$), determined by the transient spin temperatures, between $M^\uparrow$ states and $M^\downarrow$ states at four time delays ($t\ ps$) are illustrated for both cases.

**Acknowledgments:**

This work is supported by National Key Research and Development Program of China (Grant No. 2016YFA0300803), the National Natural Science Foundation of China (Grant No. 61427812, 11774160), the Natural Science Foundation of Jiangsu Province of China (No. BK20192006). X. Lu acknowledges supported by National Science Foundations of Jiangsu Province (Grants approval No. SBK2020041775).

**Author Contributions:**

J.W. and Y.X. conceived and supervised the project. G.L. performed the experiments with the help of X.Z. J.C. prepared the sample. J.L.W. and T. O. did the simulations. All contributed to the analysis of the experiment data. G.L. and J.W wrote the paper with the input from all authors.

**Competing interests:**

The authors declare no conflicts of interests.

**Data available:**

All data files are available by requesting the corresponding authors.